\documentclass[12pt]{article}
\usepackage{graphicx}
\def\be{\begin{equation}}
\def\ee#1{\label{#1}\end{equation}}
\def\mc{\mathcal}
\def\ci{\cite}
\def\ra{\rightarrow}
\def\GN{\Gamma_N}
\def\uN{_N}
\def\tot{{\mbox{\scriptsize TOT}}}
\def\max{{\mbox{\scriptsize max}}}
\def\r{{\mbox{\scriptsize R}}}
\def\i{{\mbox{\scriptsize I}}}

\begin{document}
\title{\hspace{75mm} 
{\small Nonlin. Phen. Compl. Syst. 17 (2014) 433-438}\\
\vspace{15mm}
Complex Masses of Resonances in the~Potential Approach}
\author{M. N. Sergeenko\\
{\small\itshape Gomel State Medical University}\\
5 Lange Street, Gomel, BY-246050, Belarus} 
\maketitle

\begin{abstract}
Quarkonium resonances in the~complex-mass scale are studied. 
Relativistic quark potential model is used to describe the~quark-antiquark 
system. 
The~complex-mass formula is obtained from two exact asymptotic solutions for 
the~QCD motivated potential with the~distance-dependent value of the~strong 
coupling in~QCD. 
The~centered masses and total widths of some meson resonances are calculated. 
A~possible origin of the~``dark matter'' and the~``Missing Mass''is 
discussed. 
\begin{description}
\item[PACS numbers]11.10.St; 12.39.Pn; 12.40.Nn; 12.40.Yx.
\item[Keywords]
meson,quark,bound state,resonance,complex mass,width. 
\end{description}
\end{abstract}
\maketitle

\section*{Introduction}
\label{intro}
The~concept of resonance arises from the~study of oscillating systems 
in classical mechanics and extends its applications to physical theories 
like electromagnetism, optics, acoustics, and Quantum Mechanics. 
In QM and Quantum Field Theory resonances may appear in similar 
circumstances to~classical physics. 
However, they can also be thought of as unstable particles with 
the~particle's complex-energy poles in the~scattering amplitude. 
In the~version of QFT, the~resonances are described by the~complex-mass 
poles of the~scattering matrix. 

Most particles listed in the~Particle Data Group tables~\ci{PDG2012} are 
unstable. 
A thorough understanding of the~physics summarized by the~PDG is related 
to the~concept of a~resonance. 
There are great amount and variety of experimental data and the~different 
approaches used to extract the~intrinsic properties of the~resonances. 
There is the~lack of a precise definition of what is meant by mass and 
width of resonance. 
There are two well-known definitions of mass and width of a~given 
resonance, both widely used in the~hadron physics~\ci{BerniCaPe}. 
One definition, known as the~conventional approach is based on 
the~behavior of the~{\it phase shift} $\delta$ of the~resonance as 
a~function of the~energy, while the~other, known as 
the~{\it pole approach}, is based on the~pole position of the~resonance 
and includes several approaches~\ci{MorgPenn}. 

In this work, in contrast to the~usual analysis of the~scattering 
amplitude, we consider meson resonances in bound state region, 
i.\,e., quasi-bound states of the~meson constituents. 
In traditional approach to investigate resonances one deals with 
the~scattering theory, exploring the~properties of S-matrix and 
partial amplitudes.
In contrast to the~usual analysis, meson resonances are considered to be 
the~transient oscillations of the~quark-antiquark system. 
We consider the~bound-state problem using the~potential approach and analyze 
the~mass spectrum generated from solution of the~relativistic quasi-classical 
(QC) wave equation. 

\section{The~complex-mass scale}
\label{CxMas}
The~complex numbers generalize the~real numbers. 
These numbers are important even if one wants to find real solutions 
of a~problem. 
Using complex numbers, we are getting more than what we insert. 
Remind the~important properties of complex numbers such as 
the~{\it fundamental theorem of algebra}, i.\,e., the~existence of 
$n$~roots of any $n$-th order polynomial with complex coefficients. 
As known, it would't work if we demanded real solutions. 
Holomorphic (natural) functions of a~complex variable have many important 
mathematical properties that turn complex numbers into useful if not 
essential tools, e.\,g. in the~case of two-dimensional conformal field 
theories (CFT). 
In many of the~applications, the~complex numbers may be viewed as 
non-essential but very useful technical tricks.

Operators in QM are Hermitian and the~corresponding eigenvalues are real. 
However, in scattering experiment, the~wave function requires different 
boundary condition, that is why the~complex energy is 
required~\ci{MoisPR98,Taylor}. 
There is the~connection between the~widths of resonances, $\Gamma_n$, and 
the~imaginary part of the~complex poles of the~S-matrix. 
The~full-width half-maximum of the~$n$th Lorentzian peak, 
$\Gamma_n=1/\tau_n$, is the~inverse lifetime of the~resonance state, 
$\Gamma_n=1/\tau_n$. 
This analysis shows that the~resonance phenomenon as obtained in 
a~scattering experiment is mainly controlled by poles of 
the~scattering matrix. 
They are associated with the~scattering particle and the~target which 
create an~unstable intermediate (short-living bound) state of the~system. 
This means we can start with the~bound state problem and make the~analytic 
continuation to the~scattering region.

The~concept of a~purely outgoing wave belonging to the~complex eigenvalue, 
${\mc E}_n$, was introduced in 1939 by Siegert~\ci{Sieg1939} that is 
an~appropriate tool in the~studying of resonances. 
The~resonance poles are complex eigenvalues of the~Hamiltonian,
\be
\hat\mc H(r)\Psi_n^{res}(r)=\mc E_n\Psi_n^{res}(r),\quad 
\mc E_n=M_n-i\frac{\Gamma_n}2,
\ee{ResEq}
where $\mc E_n$ is the~resonance position above the~threshold. 
Equation (\ref{ResEq}) is the~basic equation in the~resonance theory for 
time-independent Hamiltonians. 
The~resonances in scattering experiments are associated with complex 
eigenvalues of the~(unscaled) Hamiltonian which describes the~physical 
system. 
Here $\Gamma_n$ corresponds to the~inverse of resonance's lifetime, 
$\Gamma_n=1/\tau_n$, of the~resonance state. 

Resonances in QFT are described by the~complex-mass poles of the~scattering 
matrix~\ci{Taylor}. 
The~complex eigenvalue also corresponds to a~first-order pole of 
the~S-matrix~\ci{Heit1947}. 
The~masses of the~states develop imaginary masses from loop 
corrections~\ci{BauDGST,CxMasDeDi}. 
In this case, the~probability density comes from the~particle's propagator 
with the~complex mass. 

Resonances in hadron physics are complex values and can be described by 
the~complex numbers. 
Resonance is present as {\em transient oscillation} associated with 
metastable states. 
In the~pole approach the~parameters $M_p$ and $\Gamma_p$ of the~resonance 
are defined in terms of the~pole position, $s_p$, in the~complex 
Mandelshtam's $s$-plane as~\ci{BerniCaPe,MorgPenn,Hislop,BruUr} 
\be 
s_p=M_p^2-iM_p\Gamma_p.
\ee{ResPol}
Fundamentals of scattering theory and strict mathematical definition 
of resonances in QM was considered in~\ci{Hislop,BruUr}. 
The~rigorous QM definition of a~resonance requires determining the~pole 
position in the~second Riemann sheet of the~analytically continued 
partial-wave scattering amplitude in the~complex Mandelstam $s$ variable 
plane~\ci{NieArri}. 
This definition has the~advantage of being quite universal regarding 
the~pole position, but can only be applied if the~amplitude can be 
analytically continued in a~reliable way. 
It is easy to see that (\ref{ResPol}) is the~approximation of the~complex 
expression 
\be
\mc M_N^2=\left(M_N-i\frac\GN 2 \right)^2,
\ee{CxPolMaGa}
which turns in (\ref{ResPol}) when $\GN\ll M_N$, that is usually observed 
in practice. 
The~poles given by (\ref{CxPolMaGa}) are located in the~fourth quadrant of 
the~complex surface $s=\mc M^2$. 
The~complex mass eigenvalues (eigenmasses), $\mc M_N$, can be found in 
the~framework of the~quark potential model from solution of the~QC wave 
equation for the~QCD-motivated potential. 

\section{The~Cornell potential}
\label{CorPotenCxMas}
The~Cornell potential is a~special in hadron physics. 
It is fixed in an~extremely simple manner in terms of very small number 
of parameters. 
This potential is unique in that sense, if considered in the~complex-mass 
scheme, it yields the~{\it complex} eigenmasses for hadrons and resonances. 

It is known that in perturbative QCD, as in QED the~essential interaction 
at small distances is instantaneous Coulomb one-gluon exchange (OGE); 
in QCD, it is $qq$, $qg$, or $gg$ Coulomb scattering~\ci{Bjor}. 
Therefore, one expects from OGE a Coulomb-like contribution to 
the~potential, i.\,e., $V_S(r)\propto-\alpha_s/r$ at $r\rightarrow 0$. 
For large distances, in order to be able to describe confinement, 
the~potential has to~rise to~infinity. 
From lattice-gauge-theory computations~\ci{LattV} follows that this rise 
is an~approximately linear, i.\,e., $V_L(r)\simeq\sigma r+$const for 
large~$r$, where $\sigma\simeq 0.15$\,GeV$^2$ is the~string tension. 
These two contributions by simple summation lead to the~famous Cornell 
$q\bar q$ potential~\ci{LattV,EichGMR},
\be
V(r)=V_S(r)+V_L(r)\equiv-\frac 43\frac{\alpha_s}r +\sigma r;
\ee{Vcor}
its parameters are directly related to basic physical quantities 
noted above. 
The~potential (\ref{Vcor}) is one of the~most popular in hadron physics; 
it incorporates in clear form the~basic features of the~strong interaction. 
All phenomenologically acceptable QCD-inspired potentials are only 
variations around this potential. 

In hadron physics, the~nature of the~potential is very important. 
There are normalizable solutions for scalar like potentials, but not for 
vector like. 
The~effective interaction has to~be Lorentz-scalar in order to~confine 
quarks and gluons~\ci{Sucher95,SemayCeu93}. 
In our consideration, we take the~potential (\ref{Vcor}) to be 
Lorentz-scalar. 

\section{Two asymptotic solutions}
\label{TwoAsy}
Our aim is to~find the~analytic form for the~resonances' eigenmasses. 
This problem is not easy if one uses known relativistic wave equations 
with the~potential~(\ref{Vcor}). 
However, it can be done from solution of the~QC wave equation for 
this potential~\ci{SeMPL97,SePRA}. 
An~important feature of this equation is that, for two and more 
turning-point problems, it can be solved exactly by the~conventional 
WKB method~\ci{SeMPL97,SePRA}. 
We use the~advantage of analyzing the~system in the~complex-mass 
scale that has important features such as a~simpler and more general 
framework~\ci{BauDGST,CxMasDeDi}. 

To show that, analyze the~eigenvalues obtained separately for the~two 
components of the~potential (\ref{Vcor}), i.\,e., the~Coulomb term 
$V_S(r)$ and the~linear one $V_L(r)$. 
Then, using the~two-point Pad\'e approximant, we join these two exact 
asymptotic solutions; this results in the~interpolating mass 
formula~\ci{SeZC94,SeYF93}: 
\be 
M_N^2=4\left[2\sigma\tilde N-\left(\frac{\tilde\alpha m}N\right)^2 
+m^2-2\tilde\alpha\sigma\right],
\ee{Mn2int} 
where $\tilde\alpha=4\alpha_S/3$, $\tilde N=N+n_r+1/2$, 
$N=n_r+J+1$ and $m$ is the~constituent quark mass. 
The~simple mass formula (\ref{Mn2int}) describes equally well the~mass 
spectra of all $q\bar q$ and $Q\bar Q$ quarkonia ranging from the~$u\bar d$ 
($d\bar d$, $u\bar u$, $s\bar s$) states up to~the~heaviest known $Q\bar Q$ 
systems~\ci{SeZC94,SeYF93}. 
The~universal formula (\ref{Mn2int}) has been used to calculate 
the~glueball masses and Regge trajectories including 
the~Pomeron~\ci{SeEPJC12,SeEPL10}. 
It appears to be successful in many other applications. 

The~``saturating'' Regge trajectories obtained from 
(\ref{Mn2int})~\ci{SeZC94,SeYF93} were applied with success to 
the~photoproduction of vector mesons that provide an~excellent simultaneous 
description of the~high and low $-t$ behavior of 
the~$\gamma\,p\rightarrow p\,\rho$, $\omega$, 
$\phi$ cross sections~\ci{LagetPRD04,CLAS_PRL03}, given an~appropriate 
choice of the~relevant coupling constants~\ci{CLAS_PRL01,CLAS_EPJA05}. 
It was shown that the~hard-scattering mechanism is incorporated in 
an~effective way by using the~``saturated'' Regge trajectories that are 
independent of $t$ at~large $-t$~\ci{SeZC94,SeYF93}. 

The~mass formula (\ref{Mn2int}) is very transparent physically, as 
well as the~potential~(\ref{Vcor}) (Coulomb + linear). 
This formula contains in a~hidden form some important information; 
we can get it in the~following way. 
It is easy to see that (\ref{Mn2int}) is the~real part of the~complex 
expression,
\be 
\mc M_N^2=4\left[\left(\sqrt{2\sigma\tilde N}
-i\frac{\tilde\alpha m}N\right)^2
+\left(m+i\sqrt{2\tilde\alpha\sigma}\right)^2\right]
\equiv 4\left[\pi_{\uN}^2+\mu^2\right]. 
\ee{E2nCx}
This expression has the~form of equation for two free relativistic 
particles with the~complex momentum, 
\be
\pi_{\uN} =\sqrt{2\sigma\tilde N}-i\frac{\tilde\alpha m}N, 
\ee{CxPi}
and mass, 
\be
\mu=m+i\mu_{\i},\quad \mu_{\i}=\sqrt{2\tilde\alpha\sigma}.
\ee{CxQms}
The~complex-mass expression (\ref{E2nCx}) contains additional 
information, but let us give some ground to~our consideration. 

An important hint we observe studying the~hydrogen atom problem. 
The~total energy eigenvalues for the~non-relativistic Coulomb 
problem can be written with the~use of complex quantities in 
the~form of the~kinetic energy for a~{\it free} particle, 
\be
E_n=\frac{p_n^2}{2m},\quad p_n=\frac{i\alpha m}{n_r+l+1}\equiv mv_n,
\ee{Etokin} 
where $p_n$ is the~electron's momentum eigenvalue with the~imaginary 
discrete velocity, $v_n=i\alpha/(n_r+l+1)$. 
This means, that the~motion of the~electron in a~hydrogen atom is free, 
but restricted by the~``walls'' of the~potential~\ci{SeMPL97,SePRA}. 

The~relativistic two-body problem with the~scalar-like Coulomb potential 
for particles of equal masses can be~solved analytically~\ci{SeMPL97}. 
The~exact expression for the~c.\,m. energy squared is written in the~form 
of two {\it free} relativistic particles as~\ci{SeEPJC12,SeEPL10}:
\be
E_N^2=4\left[\left(i{\rm Im}\{\pi_{\uN}\}\right)^2+m^2\right],
\quad{\rm Im}\{\pi_{\uN}\}=\frac{\tilde\alpha m}N\equiv mv_{\uN},
\ee{En2Cou}
where $N$ is given above, and we have introduced the~{\it imaginary} 
momentum eigenvalues, $i{\rm Im}\{\pi_{\uN}\}$, and discrete 
velocity~$v_{\uN}$. 

The~linear term of the~Cornell potential~(\ref{Vcor}) can be dealt 
with analogously. 
In this case the~exact solution is also well 
known~\ci{SeMPL97,SeEPJC12,SeEPL10}:
\be
E_N^2=8\sigma\tilde N,\quad\tilde N=N +\left(n_r+\frac 12\right).
\ee{E2li}
This expression does not contain the~mass term and can be written 
in the~similar to~(\ref{En2Cou}) relativistic form,
\be
E_N^2=4\left({\rm Re}\{\pi_{\uN}\}\right)^2,\quad
{\rm Re}\{\pi_{\uN}\}=\sqrt{2\sigma\tilde N},
\ee{E2Re}
where ${\rm Re}\{\pi_{\uN}\}$ is the~{\it real} momentum eigenvalue. 

Thus, two asymptotic additive terms of the~potential (\ref{Vcor}), 
$V_S(r)$ and $V_L(r)$, separately, yield the~imaginary (\ref{En2Cou}) 
and real (\ref{E2Re}) momentum eigenvalues. 
These terms of the~potential represent two ``different physics'' 
(Coulomb OGE at small distances $r$ and the linear string tension at 
large $r$), therefore, two different realms of the~interaction. 
Each of these two expressions, (\ref{En2Cou}) and (\ref{E2Re}), is exact 
and was obtained independently, therefore, we can consider the~complex 
sum, $\pi_{\uN}={\rm Re}\{\pi_{\uN}\}+i{\rm Im}\{\pi_{\uN}\}$ given 
by~(\ref{CxPi}). 
Here we accept the~quark complex eigenmomentum, $\pi_{\uN}$, that means 
the~resonance total energy and mass should be complex as well. 

It is an experimental fact that the~dependence $M_N^2(J)$ is linear 
for light mesons~\ci{Collin}. 
However, at present, the~best way to reproduce the~experimental masses 
of particles is to~rescale the~entire spectrum given by~(\ref{E2li}) 
assuming that the~masses of the~mesons are expressed by 
the~relation~\ci{Sucher95}
\be 
M_N^2 = E_N^2 - C^2,
\ee{Mn2C}
where $C$ is a constant energy (shift parameter). 
Relation (\ref{Mn2C}) is used to~shift the~spectra and appears as a~means 
to~ simulate the~effects of unknown structure approximately. 
But, if we rewrite (\ref{Mn2C}) in the~usual relativistic form,
\be 
M_N^2=4\left[\left({\rm Re}\{\pi_{\uN}\}\right)^2+(\pm i\mu_{\i})^2\right],
\ee{MnImM}
where ${\rm Re}\{\pi_{\uN}\}$ is given by (\ref{E2Re}), we come to 
the~concept of the~imaginary mass, $\mu_{\i}$. 
Here in (\ref{MnImM}) we have introduced the~notation, 
$4(\pm i\mu_{\i})^2=-C^2$. 
What is the~mass $\mu_{\i}$?

The~required shift of the~spectra naturally follows from the~asymptotic 
solution of the~QC wave equation for the~potential 
(\ref{Vcor})~\ci{SeMPL97,SeEPJC12,SeEPL10,SeIMPA03}. 
To show that, we account for the~``weak coupling effect'', i.\,e., together 
with the~linear dependence in (\ref{E2Re}) we should include 
the~contribution of the~Coulomb term, $-\alpha_S/r$, of the~potential. 
These kind of calculations result in the~asymptotic expression similar 
to (\ref{Mn2C})~\ci{SeMPL97,SeEPJC12}. 
Comparing (\ref{En2alf}) with~(\ref{MnImM}), we come to the~equation:
\be
M_N^2=8\sigma(\tilde N -\tilde\alpha)\equiv 
4\left[2\sigma\tilde N + (i\mu_{\i})^2\right],
\ee{En2alf}
where $\mu_{\i}$ is the~quark imaginary-part mass. 
We see that the~additional term, $-8\tilde\alpha\sigma$ arises from 
the~interference of the~Coulomb and linear components of the~Cornell 
potential (\ref{Vcor}). 
Therefore, we have the~quark real-part (constituent) mass, $\mu_\r=m$, 
and the~imaginary-part mass (\ref{CxQms}). 
As in case of the~eigenmomenta, we introduce the~quark {\it complex} 
mass, $\mu=\mu_\r+i\mu_{\i}$ given by (\ref{CxQms}). 

The~interference term $-8\tilde\alpha\sigma=4(i\mu_{\i})^2$ in~(\ref{En2alf}) 
contains only the~parameters of the~potential (\ref{Vcor}) and is 
Lorentz-scalar, i.\,e., additive to the~particle masses. 
This is why, we accept the~last additive term in (\ref{En2alf}) to be 
the~mass term, which contributes to the~quark complex mass (\ref{CxQms}). 
This term is negative and effectively this reduces the~quark effective 
mass~(\ref{CxQms}). 
This means, that a~part of the~quark mass goes into the~interaction's 
field that causes the~quark's mass defect, which is similar to one 
we observe in nucleus (nuclear energy). 

\section{The resonance's mass and total width}
\label{ResMaWid}
Fundamentals of scattering theory and strict mathematical definition 
of resonances in QM was considered in~\ci{Hislop}. 
It is possible to extend the~Hamiltonian that describes a~quantum system 
into the~complex domain while still retaining the~fundamental properties 
of a~quantum theory. 
One of such approaches is Complex Quantum Mechanics (CQM)~\ci{BendBH}. 
The~CQM has proved to be so interesting that research activity in this 
area has motivated studies of {\it complex classical 
mechanics}~\ci{DorDuTa}. 
Complex energies and masses cannot be measured experimentally nor 
simulated by lattice QCD calculations, and basically an~extrapolation 
is needed, which is a~potentially uncontrolled arbitrary procedure. 

The~complex eigenvalues correspond to a~first-order poles of 
the~S-matrix~\ci{Heit1947}. 
In order to deduce these poles reliably, one must either have narrow 
resonances, small backgrounds, or accurate amplitudes, requirements which 
are rarely met in the~PDG compilation~\ci{PDG2012}. 
Each unstable particle in CQM is associated with a~well-defined object, 
which is a~discrete generalized eigenstate of the~Hamiltonian having its 
real and negative imaginary parts being the~centered mass, $M_N$, and 
half-width, $\Gamma_n/2$, of the~particle, respectively~\ci{BendBH,DorDuTa}. 

We have obtained two exact asymptotic solutions, i.\,e., (\ref{En2Cou}) 
and~(\ref{En2alf}). 
The~complex mass (\ref{E2nCx}) can be written in the~form (\ref{ResPol}),
\be 
\mc M_N^2={\rm Re}\{\mc M_N^2\}+i{\rm Im}\{\mc M_N^2\},
\ee{E2nReIm}
where ${\rm Re}\{\mc M_N^2\}$ is given by the~interpolating mass 
formula~(\ref{Mn2int})~\ci{SeZC94,SeYF93,SeEPJC12,SeEPL10}, 
and the~imaginary part is given by expression: 
\be 
{\rm Im}\{\mc M_N^2\}=8m\sqrt{2\tilde\alpha\sigma}
\left(1-\frac{\sqrt{\tilde\alpha\tilde N}}N\right).
\ee{ImE2n}
Comparing (\ref{E2nReIm}) and (\ref{ResPol}), we obtain the~centered 
mass squared, $M_N^2$, given by (\ref{Mn2int}), and the~total 
width, 
\be 
\GN^\tot=\sqrt{2\tilde\alpha\sigma}\frac{8m}{M_N}
\left|1-\frac{\sqrt{\tilde\alpha\tilde N}}N\right|. 
\ee{Gamn}

One can consider another resonance's definition in the~pole approach. 
In general (mathematically), the~S-matrix is a~meromorphic function 
of complex variable $\mc M=\pm\sqrt s$, where the~complex $s$-plane 
is replaced by the~two-sheet Riemann surface, $\pm\sqrt s$, made up 
of two sheets $R_0$ and $R_1$, each cut along the~positive real axis, 
Re$\mc M$, and with $R_1$ placed in front of $R_0$~\ci{Hislop,NieArri}. 
\begin{figure}[th]
\begin{center}
\includegraphics[scale=1.0]{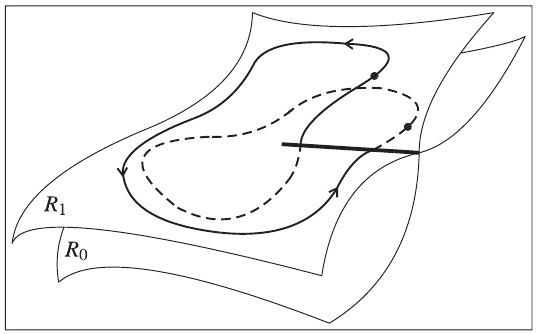}
\caption{{\small The two-sheet Riemann surface $\mc{M}=\pm\sqrt s$. 
The~lower edge of the~slit in $R_0$ is joined to the~upper edge of the~slit 
in $R_1$, and the~lower edge of the~slit in~$R_1$ is joined to the~upper 
edge of the~slit in $R_0$.}} \label{fig1:Riem2Shee}
\end{center}
\end{figure}
The resonance positions are symmetrically located in the Riemann 
$\mc{M}$-surface (Fig.\,\ref{fig1:Riem2Shee}).  

The~square root of the~complex expression (\ref{E2nReIm}) gives 
\be
\sqrt{\mc M_N^2}
=\pm\Bigl({\rm Re}\{\mc M_N\}+i\xi{\rm Im}\{\mc M_N\}\Bigr),
\ee{sqrtM2}
where 
\be
{\rm Re}\{\mc M_N\}=\sqrt{\frac{|\mc M_N^2|+{\rm Re}\{\mc M_N^2\}}2},
\ee{ReMN}
\be
{\rm Im}\{\mc M_N\}=\sqrt{\frac{|\mc M_N^2|-{\rm Re}\{\mc M_N^2\}}2}
=-\frac{\GN^\tot}2,
\ee{ImMN}
$|\mc M_N^2|=\Bigl[\Bigl({\rm Re}\{\mc M_N^2\}\Bigr)^2+
\Bigl({\rm Im}\{\mc M_N^2\}\Bigr)^2\Bigr]^{1/2}$, 
$\xi={\rm sgn}({\rm Im}\{\mc M_N^2\})$. 
The~expressions (\ref{ReMN}) and (\ref{ImMN}) define the~resonance 
position in the~Riemann $\mc M$-surface. 
The~centered mass, $M_N^R={\rm Re}\{\mc M_N\}$, and the~total width, 
$\GN^\tot=-2{\rm Im}\{\mc M_N\}$, are process independent parameters of 
the~resonance. 

Resonance masses arise in complex conjugate pairs. 
Poles in the~left half-plane correspond to either bound or anti-bound 
states~\ci{MoisPR98,LandLif}. 
If $\mc M_N=M_N-i\GN/2$ is a~pole in the~fourth quadrant of the~surface 
$\pm\sqrt s$, then $\mc M_N=-M_N-i\GN/2$ is also a~pole, but in the~third 
quadrant (antiparticle)~\ci{MoisPR98}. 
Comprehensive definitions of the resonance's parameters, i.\,e., their 
masses and widths, require further investigations. 
An~alternate definition of the~resonance's width can be obtained 
from~(\ref{ImE2n}). 
According to the~definition (\ref{ResPol}) the~width is given by 
the~{\it imaginary-part} mass of the~resonance's complex mass,~$\mc M_N$. 
Dividing~(\ref{ImE2n}) by $8m$ (exclusion of the~real-mass term), 
we come to the~following expression: 
\be
\GN^\tot=\sqrt{2\tilde\alpha\sigma}
\left|1-\frac{\sqrt{\tilde\alpha\tilde N}}N\right|. 
\ee{GamnV}
This total width is restricted by the~maximum possible value 
$\GN^\max=\mu_{\i}\equiv\sqrt{2\tilde\alpha\sigma}$ for highest excitations 
(resonances) at $N\ra\infty$. 

\section{Results and discussions}
\label{ResulDisc}
Quantum theory of resonances, calculating energies, width and cross 
sections was considered in~\ci{MoisPR98} by complex scaling. This method 
enables one to associate the~resonance phenomenon with a~single square 
integrable eigenfunction of the~complex-scaled Hamiltonian, rather than 
with a~collection of continuum eigenstates of the~unscaled Hermitian 
Hamiltonian. 

As an~example, we have considered the~$\rho$-family $J=l+1$ resonances of 
the~leading Regge trajectory, $\alpha_\rho(s)$. 
Calculation results are in Table 1, where masses and widths are in~MeV.
\begin{table}[tbh]
\begin{center}
\caption{Masses and total widths of the~$\rho$-family resonances
\label{tab:tab1}}
\begin{tabular}{llccccc}
\hline
\textrm{Meson} &$\ J^{PC}$ & $\ \ M_N^{ex}$ &$\ \ M_N^{th}$& 
$\ \ \GN^{ex}$ &$\GN^{th}(\ref{ImMN})$ &$\GN^{th}(\ref{GamnV})$\\
\hline\hline
$\rho\ (1S)$&$\ \ 1^{--}$&\ \ 776 &\ 775 & 149 & 150 & 75\\
$a_2(1P)$ &$\ \ 2^{++}$ &\ \ 1318 &\ 1323 & 107 & 108 & 93\\
$\rho_3(1D)$&$\ \ 3^{--}$&\ \ 1689 &\ 1689 & 161 & 170 & 188\\
$a_4(1F)$&$\ \ 4^{++}$ &\ \ 1996 &\ 1985 & 255 & 194 & 249\\
$\rho_5(1G)$&$\ \ 5^{--}$&\ \ $-$ &\ 2234 & $-$ & 202 & 294\\
$a_6(1H)$&$\ \ 6^{++}$ &\ \ $-$&\ 2462 & $-$ & 205 & 328\\
\hline
\end{tabular}
\end{center}
\end{table}
Parameters in these calculations are found from the~best fit to 
the~available data~\ci{PDG2012}: $\alpha_S=1.463$, string tension 
$\sigma=0.134$\,GeV$^2$, the~quark effective mass $m=193$\,MeV. 
The~widths $\GN^{th}(\ref{ImMN})$ and $\GN^{th}(\ref{GamnV})$ are 
calculated with the~use of the formulas (\ref{ImMN}) and (\ref{GamnV}). 
The~maximum width $\GN^\max=\mu_{\i}=723$\,MeV. 
More accurate calculations require taking into account the spin-spin 
and spin-orbit interactions to the~potential~(\ref{Vcor}); 
the~spin corrections have been considered in~\ci{SeZC94}. 

\begin{figure}[th]
\begin{center}
\includegraphics[scale=0.35]{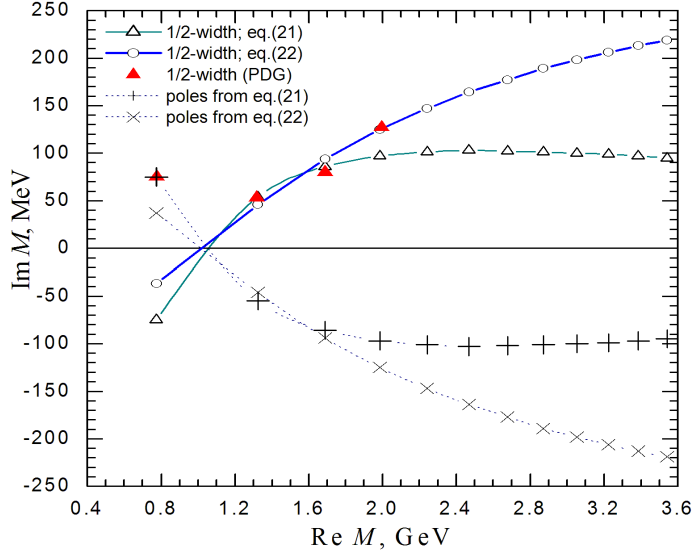}
\caption{The~complex Riemann $\mc M=\pm\sqrt s$-surface. 
Stars show location of the~complex-mass resonances relating to 
the~leading $\rho$ Regge trajectory; the~imaginary-part component 
of the~resonances gives its half-width. 
Other notations are shown on the~figure. 
Calculations done with the~use of the~formulas (\ref{ImMN}) 
and (\ref{GamnV}).} \label{fig:CxMro2}
\end{center}
\end{figure}

Location of the~$\rho$-family resonances in the~complex 
$\mc M$-surface is shown in Fig.~\ref{fig:CxMro2}. 
It is known that complex poles of the~S-matrix always arise in conjugate 
pairs~\ci{LandLif,FernOrtiz}. 
Poles in the lower half-plane are complex-conjugated with zeros in upper 
half-plane. 
Zeros above and on the~real axis correspond to bound states, and resonances 
are located in the~lower half-plane. 
Purely outgoing resonant states are defined by~poles in the~second Riemann 
sheet, i.\,e., fourth quadrant of the complex 
surface~${\mc M}=\pm\sqrt s$~\ci{Hislop,NieArri}. 

Note another feature of the~$\rho$-family resonance data. 
There is a~dip~($\GN^\tot=107$\,MeV) for the~$a_2(1320)$ resonance 
(see Fig.~\ref{fig:CxMro2}). 
This dip is described in our scheme and has the~following explanation. 
In our consideration, the~resonance's pole definition is true for 
all resonances except the~first 1S~$\rho(770)$ meson state with 
the~quantum numbers $n_r=0$ and $l=0$. 
Equations (\ref{ReMN}) and (\ref{ImMN}) define the poles' location 
in the~whole two-sheet Riemann surface $\mc M=\pm\sqrt s$. 
Analysis and numerical calculations using these formulas show that 
the~imaginary-part mass, ${\rm Im}\{\mc M_N\}$, of the~$\rho(770)$ 
resonance is positive, i.\,e., the~$\rho(770)$-state is located in 
the~first quadrant of the complex $\mc M$-surface (${\rm Re}\{\mc M_N\}>0$, 
${\rm Im}\{\mc M_N\}>0$). 
All other states of the~$\rho(770)$-family Regge trajectory are embedded 
in the~fourth quadrant (${\rm Re}\{\mc M_N\}>0$, ${\rm Im}\{\mc M_N\}<0$). 

One can give the~following explanation to this phenomenon. 
If we define resonance to be an~excited state with non-zero quantum numbers 
either $l$ or $n_r$, then $\rho(770)$ meson with $l=0$ and $n_r=0$ is not 
resonance but the~bound state with parallel spins of quarks (vector 
particle). 
This is only difference between $\rho$ and $\pi$-mesons. 
In this case all resonances have non-zero quantum numbers ($l\ne 0$ or 
$n_r\ne 0$, or both) and located below of the~real axis. 
If we accept this definition, then all $(n_r+1)l(J^{PC})=$ $1S(J^{--})$ 
states are mesons if they are located in the~first quadrant of the complex 
$\mc M$-surface. 

\section*{Conclusion}
\label{Conclu}
The~complex-scaled method is the~extension of theorems and principles, 
which were originally proved in quantum mechanics for Hermitian operators, 
to non-Hermitian operators and also on the~development of 
the~complex-coordinate scattering theory. 
The~method enables the~calculation of the~energy positions, lifetimes and 
partial widths of resonance states. 
We have studied meson resonances to be the~quasi-bound eigenstates 
of two quarks interacting by the~Cornell potential. 
Using the~complex analysis, we have derived the~quarkonia complex-mass 
formula, in which the~real and imaginary parts are exact expressions. 
This approach has allowed us to simultaneously describe in the~unified 
way the~centered masses and total widths of the~$\rho$-family resonances. 

The~complex-mass scale has relation to~some non-Hermitian Hamiltonian. 
In disagreement with a~widely spread belief that the~width $\Gamma_N(M_N)$ 
of meson resonances linearly depends on their mass, we have found this 
inconsistent with an~existence of non-linear Regge trajectories. 
We have shown, that the~Regge trajectories are non-linear analytic 
functions of the invariant squared mass $s=\mc M^2$ as their widths 
obtained in this work, which are restricted or decreasing with the~resonance 
mass~$M_N$. 

Resonances represent a~very economical way in theoretical description 
of hadronic reactions at high energies. 
Such a~task is very important nowadays since a~great significance of 
the~width of heavy resonances. 
Our analysis may be important for further development of the~string model 
of hadrons and for improvement of such transport codes as the~hadron string 
dynamics by including the~finite width of heavy resonances. 

The~complex masses and energies are not observable directly but may have 
relation to the~``Missing Mass'' and ``Dark Matter''. 
We have shown in this work that the~energy, momentum and mass of particles 
and resonances are complex. 
Reggeons in the~Regge theory (Regge trajectories $\alpha(t)$) are 
the~complex functions of the~transfered momentum $t$ (negative invariant 
squared mass $s$) are the~imaginary-mass hypothetical objects. 
But they describe the~{\it real} interactions. 
The~imaginary mass may have the~same possibility to exist as the~real one. 

The~imaginary mass is contained in the~magnitude of the~complex mass, 
$|\mc M|=(\mc M_R^2+\mc M_{\i}^2)^{1/2}$, and give contribution to observables. 
"Missing Mass" can perhaps be measured, and mass or energy apparently 
vanishing from a region of space-time may be taken as an indication that 
something is leaving that region, perhaps along another perpendicular 
axis, the~imaginary one.
This would mean the~``Missing Mass'' and ``Dark Matter''\dots 

This work was done in the~framework of investigations for the~experiment 
ATLAS (LHC), code 02-0-1081-2009/2016, 
``Physical explorations at LHC'' (JINR-ATLAS). 

\bibliography{BibData}

\begin{thebibliography}{10}

\bibitem{PDG2012}
J.~Beringer.
\newblock {\em Phys. Rev. D}, 86:010001, 2012.

\bibitem{BerniCaPe}
G.L.~Castro A.~Bernicha and J.~Pestieau.
\newblock {\em Nucl. Phys. A.}, 597:623, 1996.

\bibitem{MorgPenn}
D.~Morgan and M.~R. Pennington.
\newblock {\em Phys. Rev. D}, 48:1185, 1993.

\bibitem{MoisPR98}
N.~Moiseyev.
\newblock {\em Phys. Rep.}, 302:211, 1998.

\bibitem{Taylor}
J.~R. Taylor.
\newblock {\em The Quantum Theory of Nonrelativistic Collisions}.
\newblock Dover Publications, 2006.

\bibitem{Sieg1939}
A.~J.~F. Siegert.
\newblock {\em Phys. Rev.}, 56:750, 1939.

\bibitem{Heit1947}
W.~Heitler and N.~Hu.
\newblock {\em Phys. Rev.}, 159:776, 1947.

\bibitem{BauDGST}
T.~Bauer.
\newblock {\em {\it $et\ al.$} Complex-mass scheme and resonances in EFT. The
  8th International workshop on the physics of excited nucleons: NSTAR 2011},
  volume 1432.
\newblock AIP Conf. Proc., Reading, Massachusetts, 2011.

\bibitem{CxMasDeDi}
M.~Roth A.~Denner, S.~Dittmaier and D.~Wackeroth.
\newblock {\em Nucl. Phys. B}, 560:33, 1999.

\bibitem{Hislop}
P.~D. Hislop and C.~Villegas-Blas.
\newblock Semiclassical szego limit of resonance clasters for the hydrogen atom
  stark hamiltonian.
\newblock [arXiv:math-ph/1104.4466v1], 2011.

\bibitem{BruUr}
R.~Brummelhuis and A.~Uribe.
\newblock {\em Commun. Math. Phys.}, 136:567, 1991.

\bibitem{NieArri}
J.~Nieves and E.~R. Arriola.
\newblock {\em Phys. Lett. B}, 679:449, 2009.

\bibitem{Bjor}
J.~D. Bjorken and E.~Paschos.
\newblock {\em Phys. Rev.}, 185:1975, 1969.

\bibitem{LattV}
G.~S. Bali.
\newblock {\em Phys. Rep.}, 343:1, 2001.

\bibitem{EichGMR}
H.~Mahlke E.~Eichten, S.~Godfrey and J.~L. Rosner.
\newblock {\em Rev. Mod. Phys.}, 80:1161, 2008.

\bibitem{Sucher95}
J.~Sucher.
\newblock {\em Phys. Rev. D}, 51:5965, 1995.

\bibitem{SemayCeu93}
C.~Semay and R.~Ceuleneer.
\newblock {\em Phys. Rev. D}, 48:5965, 1993.

\bibitem{SeMPL97}
M.~N. Sergeenko.
\newblock {\em Mod. Phys. Lett. A}, 12:2859, 1997.

\bibitem{SePRA}
M.~N. Sergeenko.
\newblock {\em Phys. Rev. A}, 53:3798, 1996.

\bibitem{SeZC94}
M.~N. Sergeenko.
\newblock {\em Z. Phys. C}, 64:315, 1994.

\bibitem{SeYF93}
M.~N. Sergeenko.
\newblock {\em Phys. At. Nucl.}, 56:365, 1993.

\bibitem{SeEPJC12}
M.~N. Sergeenko.
\newblock {\em Eur. Phys. J. C}, 72:2128, 2012.

\bibitem{SeEPL10}
M.~N. Sergeenko.
\newblock {\em Europhys. Lett.}, 89:11001, 2010.

\bibitem{LagetPRD04}
J.~M. Laget.
\newblock {\em Phys. Rev. D}, 70:054023, 2004.

\bibitem{CLAS_PRL03}
M.~Battaglieri.
\newblock {\em (CLAS\ Collaboration), Phys.\ Rev.\ Lett.}, 90:022002, 2003.

\bibitem{CLAS_PRL01}
M.~Battaglieri.
\newblock {\em (CLAS\ Collaboration), Phys. Rev. Lett.}, 87:172002, 2001.

\bibitem{CLAS_EPJA05}
L.~Morand.
\newblock {\em (CLAS\ Collaboration), Eur. Phys. J. A}, 24:445, 2005.

\bibitem{Collin}
P.~D.~B. Collins.
\newblock {\em An Introduction to Regge Theory and High-Energy Physics}.
\newblock England: Cambridge Univ. Press, 1977.

\bibitem{SeIMPA03}
M.~N. Sergeenko.
\newblock {\em Int. J. Mod. Phys. A}, 18:1, 2003.
\newblock [arXiv:quant-ph/0010084].

\bibitem{BendBH}
D.~C.~Brody C.~M.~Bender and D.~W. Hook.
\newblock {\em J. Phys. A}, 41:352003, 2008.

\bibitem{DorDuTa}
C.~Dunning P.~Dorey and R.~Tateo.
\newblock {\em J. Phys. A: Math. Gen.}, 40:R205, 2007.

\bibitem{LandLif}
L.~D. Landau and E.~M. Lifshitz.
\newblock {\em Quantum Mechanics}.
\newblock Pergamon, 1965.

\bibitem{FernOrtiz}
N.~Fern\'andez-Garc\'ia and O.~Rosas-Ortiz.
\newblock {\em Ann. Phys.}, 323:1397, 2008.

\end{thebibliography}
\bibliographystyle{unsrt}
\end{document}